%% file: main.tex
\documentclass[9pt,conference]{IEEEtran} 
\pdfoutput=1 
\IEEEoverridecommandlockouts

\usepackage[english]{babel}
\usepackage{cite}
\usepackage{amsmath,amssymb,amsfonts}
\usepackage{algorithm}
\usepackage{algorithmic}
\usepackage{graphicx}
\usepackage{textcomp}
\usepackage{dsfont}
\usepackage{xcolor}
\usepackage{units}
\usepackage{booktabs}
\usepackage{comment}
\usepackage{url}
\usepackage[normalem]{ulem}
\usepackage[hidelinks]{hyperref}
\usepackage{flushend}
\usepackage{lettrine}
\usepackage{lipsum}
\usepackage{accents}
\usepackage{stfloats}
\usepackage{epstopdf}
\usepackage{array}
\usepackage{tabularx}
\usepackage{threeparttable}
\usepackage{pdfpages}

\usepackage{pgfgantt}

\usepackage{eso-pic}

\newtheorem{theorem}{Theorem}[section]

\def\BibTeX{{\rm B\kern-.05em{\sc i\kern-.025em b}\kern-.08em
    T\kern-.1667em\lower.7ex\hbox{E}\kern-.125emX}}

\newif\ifmargincomments
\margincommentstrue 

\newif\ifrev
\revtrue 

\input{Tools/def}

\usepackage[utf8]{inputenc}
\usepackage[english]{babel}

\begin{document}

\title{\LARGE \textbf{Geometric Scaling of Battery Cells and Its Effect on Key Performance Indicators}}

\author{
Tim de Krijger, Jorn van Kampen, Mouhriz Boulghalgh, and Theo Hofman
\thanks{
The authors are with the Department of Mechanical Engineering,
Control Systems Technology, Eindhoven University of Technology,
Eindhoven, The Netherlands.
Corresponding author: T. Hofman
(\texttt{t.hofman@tue.nl}).
}
}

\maketitle
\thispagestyle{empty}
\pagestyle{empty}


\AddToShipoutPictureBG*{%
\AtPageUpperLeft{%
\setlength\unitlength{1in}%
\hspace*{\dimexpr0.5\paperwidth\relax}
\makebox(0,-0.8)[c]{%
\footnotesize
\begin{tabular}{c}
T. de Krijger, J. van Kampen, M. Boulghalgh, and T. Hofman,\\
``Geometric Scaling of Battery Cells and Its Effect on Key Performance Indicators,''\\
accepted for the 2026 IEEE Vehicle Power and Propulsion Conference,
Lyon, France,\\
uploaded to arXiv on \today
\end{tabular}}}}

\AddToShipoutPictureBG*{%
\AtPageUpperLeft{%
\setlength\unitlength{1in}%
\hspace*{\dimexpr0.5\paperwidth\relax}
\makebox(0,-20.65)[c]{%
\scriptsize
\begin{tabular}{c}
© 2026 IEEE. Personal use of this material is permitted. Permission from IEEE must be obtained\\
for all other uses, including reprinting or republishing this material for advertising or promotional purposes,\\
creating new collective works, resale or redistribution to servers or lists, or reuse of any copyrighted component.
\end{tabular}}}}


\begin{abstract}
This paper presents a computationally lightweight scaling model for cylindrical lithium-ion battery cells, intended for early-stage battery design-space exploration. The model maps selected geometric and electrode-level design variables, including cell height, cell diameter, cathode active loading, and cathode porosity, to cell-level performance indicators such as capacity, DC internal resistance, mass, volume, and winding length. The scaling model is validated against available cylindrical cell data by comparing predicted capacity, internal resistance, and winding length. The validated model is subsequently used in a single-cell design-space exploration and global sensitivity analysis to evaluate capacity, internal resistance, gravimetric energy density, and volumetric energy density. The results identify the dominant design variables, favourable parameter directions, and key trade-offs between cell geometry, electrode loading, resistance, and energy density. The proposed model provides a basis for future integration into higher-level battery system and vehicle optimization frameworks.
\end{abstract}

\begin{IEEEkeywords}
battery modelling, battery scaling, design-space exploration, sensitivity analysis
\end{IEEEkeywords}

\input{Sections/Introduction}
\input{Sections/Methodology}
\input{Sections/Results}
\input{Sections/Conclusion}
\input{Sections/Acknowledgment}
\clearpage

\bibliographystyle{customIEEEtran}
\bibliography{sources}
\appendices
\input{Sections/Apendix/Appendix}

\end{document}

%% file: Tools/def.tex
\ifmargincomments

\else

\fi

\ifrev

\else

\fi

\newcommand{\flexbrac}[1]{\if\relax\detokenize{#1}\relax \else (#1) \fi}
\newcommand{\flexcomma}[1]{\if\relax\detokenize{#1}\relax \else ,#1 \fi}



%% file: Sections/Introduction.tex
\section{Introduction}
The increasing demand for electrified mobility and renewable energy storage has intensified the need for improved battery system design. Lithium-ion battery cells are used in a wide range of applications, including electric vehicles, grid storage, uninterruptible power supplies, and portable devices \cite{Soltani2021}. In these applications, battery cell design choices influence system performance through their effect on capacity, internal resistance, mass, volume, and thermal behaviour.

Lithium-ion battery cells are commonly manufactured in three main form factors: cylindrical, pouch, and prismatic cells. Each form factor has specific manufacturing characteristics and can be combined with different cell chemistries. As a result, cell format and geometry influence not only packaging and manufacturability, but also electrical and thermal performance. This paper focuses on cylindrical cells, for which changes in height, diameter, and electrode-level properties affect the internal jelly-roll geometry and therefore the resulting cell performance.

\begin{figure}[t!]
    \centering
    \includegraphics[width=0.9\linewidth]{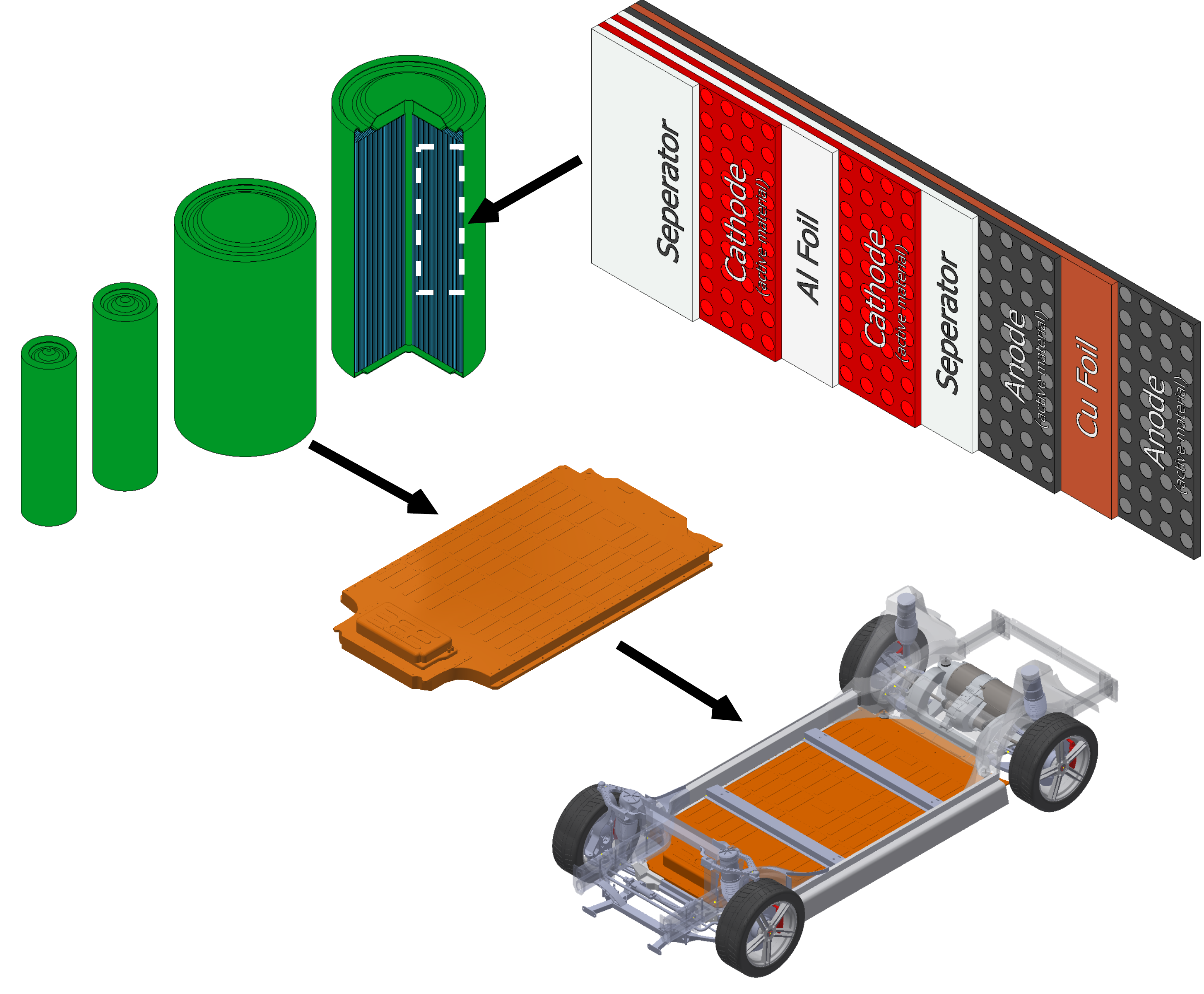}
    \caption{Overview of the geometric scaling principles used to map cylindrical cell dimensions and electrode properties to cell-level performance indicators. The battery pack and vehicle illustrate possible future integration into higher-level optimization frameworks (adapted from \cite{TeslaModelSGrabCAD}).}
    \label{fig: problemoverview}
\end{figure}

To support early-stage battery design optimization, this paper develops a simplified scaling model for cylindrical lithium-ion cells. The model maps selected geometric and electrode-level design variables to cell-level performance indicators such as capacity, DC internal resistance, mass, and volume. The proposed model is intended to provide scalable cell data for future integration into higher-level battery system and vehicle optimization frameworks, as illustrated in \autoref{fig: problemoverview}.

\subsection{Related literature}
Battery cell design has been studied from several perspectives, including electrochemical modelling, thermal behaviour, manufacturing, and cell format selection. For lithium-ion batteries, the selected cell format influences packaging, manufacturability, thermal behaviour, and electrical performance. Existing studies have therefore investigated how cell dimensions and form factor affect the performance of commercial battery cells.

Several studies compare existing cell formats and cylindrical cell designs. Comparisons between 18650 and 21700 cells, teardown studies of large-format cells such as the Tesla 4680, and size-effect studies on cylindrical cells show that cell dimensions influence electrochemical, thermal, mechanical, and geometrical properties \cite{Waldmann2020,Gorsch2025,Liu2024}. Other work focuses on tabless cylindrical cells, housing material, manufacturing choices, fast-charging performance, and cost modelling \cite{Pegel2023,Pegel2024}. These studies provide valuable insight into existing cell designs, but they are mainly based on specific commercial cells or manufacturing concepts.

Detailed electrochemical and multi-scale models provide deeper insight into cell behaviour, ageing, and parameterization \cite{doyle_modeling_1993,thorat_electrode_design_2009,chen_parameterization_2020,di_prima_calendar_aging_2025}. However, these models typically require many parameters and are less suited for fast early-stage design-space exploration. Therefore, a gap remains for a computationally lightweight scaling model that links cylindrical cell geometry and selected electrode-level design variables to key performance indicators such as capacity, internal resistance, mass, and volume. This paper addresses this gap by developing and validating such a scaling model for cell-level optimization and future integration into higher-level battery system design.

\subsection{Statement of Contributions} \label{sec: contribution}
This paper presents a computationally lightweight scaling model for cylindrical lithium-ion battery cells. The model maps a limited set of geometric and electrode-level design variables to key performance indicators such as capacity, DC internal resistance, mass, and volume.

The contribution of this work is twofold. First, the proposed model combines geometric scaling with electrode-level parameter variation in a form that is suitable for early-stage design-space exploration and optimization. Second, the model is validated against available cylindrical cell data and subsequently used to identify design trends for capacity, resistance, gravimetric energy density, and volumetric energy density. These results provide a basis for future integration of scalable cell models into cell-to-vehicle optimization frameworks.

\subsection{Organization}
The remainder of this paper is structured as follows. Section~\ref{sec: Methodology} presents the proposed cylindrical cell scaling model and explains how the selected design variables are mapped to cell capacity, internal resistance, mass, and volume. This section also introduces the single-cell design-space exploration and sensitivity analysis used to evaluate the different cell-level performance indicators. Section~\ref{sec: Results} presents the validation of the scaling model using benchmark cylindrical cells and discusses the resulting design trends and trade-offs. Finally, Section~\ref{sec: conclusion} concludes the paper by summarizing the main findings and outlining directions for future work.

%% file: Sections/Methodology.tex
\section{Methodology}\label{sec: Methodology}
This section presents the methodology used to map cylindrical cell design variables to cell-level performance indicators. The approach consists of two main parts. First, a scaling model is formulated that relates the selected cell design variables to the internal geometry of the cylindrical cell, including the coated electrode area and winding length. Based on this scaled geometry, the model calculates the cell capacity, DC internal resistance, mass, and volume. Second, the scaling model is used in a single-cell design-space exploration and global sensitivity analysis to evaluate the influence of the design variables on capacity, internal resistance, gravimetric energy density, and volumetric energy density.

The remainder of this section is structured as follows. Section~\ref{sec: Scaling Model} introduces the scaling model and the selected design variables. The capacity and resistance relations are then derived from the scaled cell geometry and fixed chemistry-dependent parameters. Section~\ref{sec:single_cell_analysis} describes the single-cell design-space exploration and sensitivity analysis used to evaluate the different performance indicators.

\subsection{Scaling Model}\label{sec: Scaling Model}
The scaling model describes the relationship between the cell's geometric properties and its electrical properties, namely the capacity $Q_0$ and the DC internal resistance (DCIR) $R_{\mathrm{cell}}$. Furthermore, the model relates the geometric properties to the cell mass. The model is parameterized by four independent design variables, collected in the design vector

\par\nobreak\vspace{-5pt}
\begingroup
\allowdisplaybreaks
\begin{small}
\begin{equation}
\mathbf{x}
=
\begin{bmatrix}
H_{\mathrm{cell}} &
D_{\mathrm{cell}} &
\sigma_{\mathrm{act,cat}} &
\varepsilon_{\mathrm{cat}}
\end{bmatrix}^{\mathrm{T}},
\label{eq:design_vector}
\end{equation}
\end{small}%
\endgroup

where $H_{\mathrm{cell}}$ is the cell height, $D_{\mathrm{cell}}$ is the cell diameter, $\sigma_{\mathrm{act,cat}}$ is the cathode active loading, defined as the mass of active material per unit electrode area, and $\varepsilon_{\mathrm{cat}}$ is the cathode porosity. Porosity represents the fraction of pore volume within the porous electrode.

The anode porosity $\varepsilon_{\mathrm{an}}$ is not treated as an independent design variable. To reduce the number of design variables and maintain feasible electrode combinations, it is expressed as a function of the cathode porosity. This relation reflects the assumption that a relatively porous cathode is combined with a relatively dense anode, or vice versa \cite{chen_parameterization_2020,doyle_modeling_1993}. A linear mapping is used:

\par\nobreak\vspace{-5pt}
\begingroup
\allowdisplaybreaks
\begin{small}
\begin{equation}
\varepsilon_{\mathrm{an}}
=
\varepsilon_{\mathrm{an,min}}
+
\frac{
\varepsilon_{\mathrm{cat}}-\varepsilon_{\mathrm{cat,min}}
}{
\varepsilon_{\mathrm{cat,max}}-\varepsilon_{\mathrm{cat,min}}
}
\left(
\varepsilon_{\mathrm{an,max}}-\varepsilon_{\mathrm{an,min}}
\right).
\label{eq: porosity an}
\end{equation}
\end{small}%
\endgroup
The mapping is based on the feasible ranges of the anode and cathode porosities \cite{thorat_electrode_design_2009,chen_parameterization_2020}. All design variables are bounded to ensure feasible cell designs:
\par\nobreak\vspace{-5pt}
\begingroup
\allowdisplaybreaks
\begin{small}
\begin{equation}
    \underline{\mathbf{x}} \le \mathbf{x} \le \overline{\mathbf{x}}.
\label{eq:design_variable_bounds}
\end{equation}
\end{small}%
\endgroup

where $\underline{\mathbf{x}}$ and $\overline{\mathbf{x}}$ define the minimum and maximum values for the design variables, respectively. The relation of $\mathbf{x}$ to capacity is described in the following sections, starting with the capacity relation.

\subsubsection{Capacity Relation}
The overall relation between capacity and the geometric properties can be described by
\par\nobreak\vspace{-5pt}
\begingroup
\allowdisplaybreaks
\begin{small}
\begin{equation}
    Q_{\mathrm{rev}} = Q_{\mathrm{ch,1}} \cdot \eta_{\mathrm{FC}},
    \label{eq: capacity cell}
\end{equation}
\end{small}%
\endgroup
where the initial charge capacity $Q_{\mathrm{ch,1}}$ depends on the geometric and internal variables and the First Cycle Efficiency ($\eta_{\mathrm{FC}}$), which is a chemistry-dependent cell parameter. The $\eta_{\mathrm{FC}}$ is given by
\par\nobreak\vspace{-5pt}
\begingroup
\allowdisplaybreaks
\begin{small}
\begin{equation}
\eta_{\mathrm{FC}}
= \min\!\left(
\eta_{\mathrm{FC,cat}},
\eta_{\mathrm{FC,an}}
\right),
\end{equation}
\end{small}%
\endgroup
where $\eta_{\mathrm{FC,an}}$ and $\eta_{\mathrm{FC,cat}}$ are the respective first cycle efficiencies of the anode and cathode. Both first-cycle efficiencies are chemical properties and thus parameters.
The initial charge is calculated using the following equation:
\par\nobreak\vspace{-5pt}
\begingroup
\allowdisplaybreaks
\begin{small}
\begin{equation}
    Q_{\mathrm{rev}}
= Q_{\mathrm{ch,1,cat}} \cdot m_{\mathrm{act,cat}},
\end{equation}
\end{small}%
\endgroup
where $ Q_{\mathrm{ch,1,cat}}$ is the first cycle charge of the cathode and $m_{\mathrm{act,cat}}$ is the mass of the active material of the cathode. This is defined by
\par\nobreak\vspace{-5pt}
\begingroup
\allowdisplaybreaks
\begin{small}
\begin{equation}
m_{\mathrm{act,cat}}
= \sigma_{\mathrm{act,cat}} \cdot A_{\mathrm{coat,cat}},
\end{equation}
\end{small}%
\endgroup
where $A_{\mathrm{coat,cat}}$ is the area of the active material for the cathode, defined by

\par\nobreak\vspace{-5pt}
\begingroup
\allowdisplaybreaks
\begin{small}
\begin{equation}
A_{\mathrm{coat,cat}} = 2\cdot H_{\mathrm{coat,cat}}\cdot L_{\mathrm{coat,cat}},
\label{eq: coated cathode area}
\end{equation}
\end{small}%
\endgroup
where the factor 2 follows from the fact that the current collector of the cathode is coated on both sides (see Fig.~\ref{fig: problemoverview}), $L_{\mathrm{coat,cat}}$ is the coated cathode length, $H_{\mathrm{coat,cat}}$ is the coated cathode height, defined by
\par\nobreak\vspace{-5pt}
\begingroup
\allowdisplaybreaks
\begin{small}
\begin{equation}
\begin{split}
H_{\mathrm{coat,cat}} =
H_{\mathrm{cell}}
- t_{\mathrm{cell}}
- t_{\mathrm{cap}}
-\\ H_{\mathrm{headspace}}
+ \Delta H_{\mathrm{cat}},
\end{split}
\end{equation}
\end{small}%
\endgroup
where $t_{\mathrm{cell}}$ and $t_{\mathrm{cap}}$ represent the structural thicknesses of the cell base and cap, respectively, $H_{\mathrm{headspace}}$ corresponds to the reserved internal free volume, and $\Delta H_{\mathrm{cat}}$ represents an applied height correction to account for mechanical compression, tolerances, or assembly-induced deformation. This is considered a cell property and is assumed to be constant.
The coated cathode length used in Eq.~\ref{eq: coated cathode area} is defined by
\par\nobreak\vspace{-5pt}
\begingroup
\allowdisplaybreaks
\begin{small}
\begin{equation}
L_{\mathrm{coat,cat}} =
L_{\mathrm{stack}}
- L_{\mathrm{mandrel}}
+ \Delta L_{\mathrm{cathode}},
\end{equation}
\end{small}%
\endgroup
where $L_{\mathrm{stack}}$ is the length of the wound electrode for a cylindrical cell, $L_{\mathrm{mandrel}}$ represents the length excluded by the hollow core in the center, and $\Delta L_{\mathrm{cathode}}$ denotes the coated length offset. This offset accounts for the difference between the winding length and the coated length. This is considered a cell property.
The winding length of the electrode $L_{\mathrm{stack}}$ follows from the arc length of the wound spiral\cite{Pegel2023}, and is defined as
\par\nobreak\vspace{-5pt}
\begingroup
\allowdisplaybreaks
\begin{small}
\begin{equation}
\begin{split}
L_{\mathrm{stack}}
=
\frac{t_{\mathrm{stack}}}{2} \cdot
\big[
\Theta_{\mathrm{stack}} \cdot \sqrt{1 + \Theta_{\mathrm{stack}}^{2}}
+\\
 \ln \cdot \big(
\Theta_{\mathrm{stack}} + \sqrt{1 + \Theta_{\mathrm{stack}}^{2}}
\big)\big],
\end{split}
\label{eq: stack length}
\end{equation}
\end{small}%
\endgroup
where $t_{\mathrm{stack}}$ represents the thickness of the jelly-roll, and $\Theta_{\mathrm{stack}}$ is a dimensionless winding parameter related to the outer and inner radii of the spiral geometry, defined by

\par\nobreak\vspace{-5pt}
\begingroup
\allowdisplaybreaks
\begin{small}
\begin{equation}
\Theta_{\mathrm{stack}}
=
\frac{D_{\mathrm{inner}}}{2}
\cdot
\frac{2\pi}{t_{\mathrm{stack}}},
\end{equation}
where $D_{\mathrm{inner}}$ is the internal diameter of the cell canister. The internal diameter is determined with
\begin{equation}
D_{\mathrm{inner}}
=
D_{\mathrm{cell}}
-
2\cdot t_{\mathrm{cell}}.
\end{equation}
\end{small}%
\endgroup
The jelly-roll thickness used in Eq.~\ref{eq: stack length} is dependent on $t_{\mathrm{electrode,sep}}$, $t_{\mathrm{electrode,an}}$ and $t_{\mathrm{electrode,cat}}$ which are the respective thicknesses of the separator, anode and cathode within the jelly-roll. The jelly-roll thickness and the following equations are generalized using $j$ to indicate the anode and cathode
\par\nobreak\vspace{-5pt}
\begingroup
\allowdisplaybreaks
\begin{small}
\begin{equation}
    j \in \{\mathrm{an}, \mathrm{cat}\}
\end{equation}
\end{small}%
\endgroup

\par\nobreak\vspace{-5pt}
\begingroup
\allowdisplaybreaks
\begin{small}
\begin{equation}
t_{\mathrm{stack}}
=  2 \cdot t_{\mathrm{sep}} +
\sum_{j}
 t_{\mathrm{electrode,j}}
 \label{eq: t_stack}
\end{equation}
\end{small}%
\endgroup
where $t_{\mathrm{sep}}$ is the thickness of the separator. The separator thickness is multiplied by two to take into account the separator on both sides where the anode and cathode meet within the wound jelly-roll. For the anode and the cathode, $t_{\mathrm{electrode},\mathrm{i}}$ is defined as

\par\nobreak\vspace{-5pt}
\begingroup
\allowdisplaybreaks
\begin{small}
\begin{equation}
t_{\mathrm{electrode,j}}
=
2\ t_{\mathrm{j}}
+
t_{\mathrm{foil,j}},
\label{eq: thickness_electrode_cathode}
\end{equation}
\end{small}%
\endgroup
where $t_{\mathrm{foil,j}}$ is the foil thickness of the anode and cathode, which is a cell property, and $t_{\mathrm{j}}$ is the thickness of the coated material for both the anode and the cathode. This is calculated using
\par\nobreak\vspace{-5pt}
\begingroup
\allowdisplaybreaks
\begin{small}
\begin{equation}
t_{\mathrm{j}}
=
\frac{\sigma_{\mathrm{act,j}}}{\rho_{\mathrm{act,j}}}
,
\end{equation}
\end{small}%
\endgroup
where  $\sigma_{\mathrm{act,j}}$ is the active loading for the anode and cathode, respectively, $\rho_{\mathrm{act,j}}$ denotes the density of the active material used in the cell, and is defined as
\par\nobreak\vspace{-5pt}
\begingroup
\allowdisplaybreaks
\begin{small}
\begin{equation}
\rho_{\mathrm{act,j}}
=
\rho_{\mathrm{electrode,j}}
\cdot f_{\mathrm{act,j}},
\end{equation}
\end{small}%
\endgroup
where $f_{\mathrm{act,j}}$ is the active content ratio for the anode and cathode, respectively. The active content ratio is a chemical property, $\rho_{\mathrm{electrode,j}}$ is the electrode density. It is defined by
\par\nobreak\vspace{-5pt}
\begingroup
\allowdisplaybreaks
\begin{small}
\begin{equation}
\rho_{\mathrm{electrode,j}}
=
\rho_{\mathrm{comp,j}} \cdot
\left(
1 - \varepsilon_{\mathrm{j}}
\right),
\end{equation}
\end{small}%
\endgroup
where $\rho_{\mathrm{comp,j}}$ represents the density of the electrolyte for the anode and cathode, which are chemical properties. $\varepsilon_{\mathrm{j}}$ denotes the porosity of the anode and cathode, respectively. The porosity of the cathode is used as an input, while, to limit the number of decision variables, the porosity of the anode is expressed as a function of the cathode porosity as seen in Eq.~\ref{eq: porosity an}.

The active loading of the anode, unlike that of the cathode, is not a decision variable and is defined as
\par\nobreak\vspace{-5pt}
\begingroup
\allowdisplaybreaks
\begin{small}
\begin{equation}
\sigma_{\mathrm{act,an}}
=
\frac{
Q_{\mathrm{Areal,an}}
}{
Q_{\mathrm{dis,1,an}}
}
,
\end{equation}
\end{small}%
\endgroup
where $Q_{\mathrm{Areal,an}}$ is the first cycle discharge capacity of the anode. $Q_{\mathrm{dis,1,an}}$ is the areal capacity of the anode, given by
\par\nobreak\vspace{-5pt}
\begingroup
\allowdisplaybreaks
\begin{small}
\begin{equation}
Q_{\mathrm{Areal,an}}
=
\mathrm{NPR}
\cdot
Q_{\mathrm{Areal,cat}},
\end{equation}
\end{small}%
\endgroup
where $\mathrm{NPR}$ is the negative (anode) to positive (cathode) capacity ratio. This ratio ensures the battery cell's viability in this model. It ensures that all lithium ions from the cathode can be accommodated on the anode, thus preventing metallic lithium plating. If the model uses active anode loading as an input variable, ensure that this ratio remains above 1.0. Furthermore, $Q_{\mathrm{Areal,an}}$ is the areal capacity of the anode, defined as
\par\nobreak\vspace{-5pt}
\begingroup
\allowdisplaybreaks
\begin{small}
\begin{equation}
Q_{\mathrm{Areal,cat}}
=
\sigma_{\mathrm{act,cat}}
\cdot
Q_{\mathrm{dis,1,cat}}
,
\end{equation}
\end{small}%
\endgroup
where $Q_{\mathrm{dis,1,cat}}$ is the first cycle discharge of the cathode, which is a chemical property. The mandrel length $L_{\mathrm{mandrel}}$ uses the same principle as the stack length, as per Eq.~\ref{eq: stack length}, and is defined as
\par\nobreak\vspace{-5pt}
\begingroup
\allowdisplaybreaks
\begin{small}
\begin{equation}
\begin{aligned}
L_{\mathrm{mandrel}}
&=
\frac{t_{\mathrm{stack}}}{2}
\Big[
\Theta_{\mathrm{mandrel}} \sqrt{1 + \Theta_{\mathrm{mandrel}}^{2}}
\\
&\quad
+
\ln \Big(
\Theta_{\mathrm{mandrel}} + \sqrt{1 + \Theta_{\mathrm{mandrel}}^{2}}
\Big)
\Big],
\end{aligned}
\end{equation}
\end{small}%
\endgroup
where $\Theta_{\mathrm{mandrel}}$ is the mandrel theta and is defined by
\par\nobreak\vspace{-5pt}
\begingroup
\allowdisplaybreaks
\begin{small}
\begin{equation}
\Theta_{\mathrm{mandrel}}
=
\frac{D_{\mathrm{mandrel}}}{2}
\cdot
\frac{2\pi}{t_{\mathrm{stack}}},
\end{equation}
\end{small}%
\endgroup
where $D_{\mathrm{mandrel}}$ is the mandrel diameter. The mandrel represents the diameter of the tool used for winding the jelly-roll in manufacturing. The minimum achievable bending radius of the electrode stack and the cell diameter both influence the mandrel diameter. After manufacturing, the tool is removed, and the relaxed jelly-roll retains a hollow core. In addition to its manufacturing role, the mandrel also serves a safety function in the event of thermal runaway \cite{pegel_cell_dimensions_2023}. According to \cite{pegel_cell_dimensions_2023}, the linear relation between the cell diameter and mandrel diameter is given by
\par\nobreak\vspace{-5pt}
\begingroup
\allowdisplaybreaks
\begin{small}
\begin{equation}
    D_{\mathrm{mandrel}} = D_{\mathrm{mandrel,ref}} + \frac{(D_{\mathrm{cell}}-D_{\mathrm{cell,ref}})}{20000},
\end{equation}
\end{small}%
\endgroup
where 20000 is the reported scaling factor.

\subsubsection{Resistance Relation}
 The overall cell resistance is split into three separate parts. These are classified as geometric, ionic, and electronic resistance. The geometric resistance is the resistance due to the current collectors and the tabs connecting them to both poles. The number of tabs depends on both geometric properties and manufacturing methods.

In \cite{Pegel2023}, the electronic resistance was mainly attributed to the current collector in this research. This was expanded to also account for changes in the electrolyte content of the jelly-roll, based on changes in active loading and porosity, and their effects on ionic and electronic resistance. The geometric resistance, which is in line with \cite{Pegel2023}, however, instead of using a finite element method, the jelly-roll is approximated as a plate and modeled as a single resistance, with resistance in both the radial and axial directions. The three parts are then also individually calculated for the cathode, anode, and separator, which is indicated by 
\par\nobreak\vspace{-5pt}
\begingroup
\allowdisplaybreaks
\begin{small}
\begin{equation}
    i \in \{\mathrm{an}, \mathrm{sep}, \mathrm{cat}\}.
\end{equation}
\end{small}%
\endgroup
The ionic resistance is defined as
\par\nobreak\vspace{-5pt}
\begingroup
\allowdisplaybreaks
\begin{small}
\begin{equation}
R_{\mathrm{ion},i} =
\frac{t_i}{\alpha_i \cdot \kappa_{\mathrm{eff,i}} \cdot A_i},
\end{equation}
\end{small}%
\endgroup
where $t_i$ and $\kappa_{\mathrm{eff},i}$ denote the thickness and effective ionic conductivity of layer $i$, respectively, and $A$ is the active electrode area, which, in the case of the separator, is equal to the winding area. The thickness of the individual components is derived with Eq.~\ref{eq: thickness_electrode_cathode}. The effective ionic conductivity is defined as 
\par\nobreak\vspace{-5pt}
\begingroup
\allowdisplaybreaks
\begin{small}
\begin{equation}
\kappa_{\mathrm{eff}}
=
\kappa_{\mathrm{bulk},i}
\cdot
\varepsilon_j^{\,\beta},
\end{equation}
\end{small}%
\endgroup
where $\kappa_{\mathrm{bulk}}$ is the bulk ionic conductivity of the electrolyte phase, $\varepsilon$ is the porosity of the medium, and $\beta$ is the Bruggeman exponent accounting for the effect of pore tortuosity and microstructural connectivity on ionic transport. In this research, the Bruggeman exponent is taken as 1.5 \cite{doyle_modeling_1993}. 

The total ionic resistance contribution of the stack is obtained by summing the cathode, anode, and separator contributions:

\par\nobreak\vspace{-5pt}
\begingroup
\allowdisplaybreaks
\begin{small}
\begin{equation}
R_{\mathrm{ion,stack}}
=
\sum_{i}
\frac{t_i}
{\alpha_i \cdot \kappa_{\mathrm{bulk},i} \cdot
\varepsilon_{\mathrm{i}}^{\,\beta} \cdot A_i}.
\label{eq: R_ion}
\end{equation}
\end{small}%
\endgroup

The next contribution to the DCIR of the battery cell, the electronic conductivity of the individual parts, is defined as
\par\nobreak\vspace{-5pt}
\begingroup
\allowdisplaybreaks
\begin{small}
\begin{equation}
R_{\mathrm{el},i} =
\frac{L_i}{\sigma_{\mathrm{eff},i} \cdot A_i},
\end{equation}
\end{small}%
\endgroup
where $\sigma_{\mathrm{eff}, i}$ is the effective electronic conductivity of the active material, $A_i$ is the area of the active material, and $L_i$ is the length of the coated electrode. The effective electronic conductivity of the active material phase is defined as
\par\nobreak\vspace{-5pt}
\begingroup
\allowdisplaybreaks
\begin{small}
\begin{equation}
\sigma_{\mathrm{eff},i} = \sigma_{\mathrm{bulk},i} \cdot (1 - \varepsilon_i)^{\beta},
\end{equation}
\end{small}%
\endgroup
where $\sigma_{\mathrm{bulk},i}$ is the bulk electronic conductivity of the solid phase, $\varepsilon_i$ is the porosity of electrode $i$, and $\beta$ is the Bruggeman exponent accounting for microstructural effects and electronic percolation within the porous electrode. 

The total electronic resistance contribution of the cell is obtained by summing the cathode and anode contributions:

\par\nobreak\vspace{-5pt}
\begingroup
\allowdisplaybreaks
\begin{small}
\begin{equation}
R_{\mathrm{el,cell}} =
\sum_{i}
\frac{L_i}
{\sigma_{\mathrm{bulk},i} \cdot (1 - \varepsilon_i)^{\beta} \cdot A_i}.
\label{eq: R_el}
\end{equation}
\end{small}%
\endgroup

\noindent The geometric resistance contribution of the current collectors is determined from the material resistivity and the effective current conduction path in both the radial and axial direction\cite{Pegel2023}. The radial resistance is approximated as the resistance through a plate. The jelly-roll's rolled-out form represents the plate. Using this, the radial resistance of the current collector $i$ is defined as
\par\nobreak\vspace{-5pt}
\begingroup
\allowdisplaybreaks
\begin{small}
\begin{equation}
R_{\mathrm{geom},i}^{\mathrm{circ}} =
\rho_i \cdot
\frac{L_{\mathrm{eff},i}}
{3 \cdot t_{\mathrm{foil},i}\cdot H_{\mathrm{stack}}},
\label{eq: R_geom_circ}
\end{equation}
\end{small}%
\endgroup
\noindent where $\rho_i$ is the electrical resistivity of the current collector material, $L_{\mathrm{w}, i}$ is the effective winding length, $t_{\mathrm{foil}, i}$ is the foil thickness. The literature shows that the model deviates from a simple 1-D $\frac{L}{A}$ model, but this deviation is compensated for with a geometric factor, in this paper that geometric factor is assumed to be 3. The foil thickness of the current collector is assumed to be a constant cell property \cite{taheri_constriction_resistance_2013,campillo_robles_current_collector_thickness_2017}.
\noindent For the cathode and anode, the effective winding lengths are defined as
\par\nobreak\vspace{-5pt}
\begingroup
\allowdisplaybreaks
\begin{small}
\begin{equation}
L_{\mathrm{eff},i} = \frac{L_{\mathrm{w}}}{N_{\mathrm{tab},i} \cdot 2},
\qquad
\label{eq: effective lenght}
\end{equation}
\end{small}%
\endgroup
where $N_{\mathrm{tab,i}}$ is the number of connections from the current collector to the cell poles. Within this paper, this will be referred to as the number of tabs. The number of tabs depends on the manufacturing methods, as shown in \cite{di_prima_calendar_aging_2025,batterymooch_p50b_tabless_2024,molicell_p45b_teardown_2023}. For the scaling model, the manufacturing method that yields the lowest overall resistance is selected to indicate the number of tabs as a function of the cell diameter. This relation is defined by
\par\nobreak\vspace{-5pt}
\begingroup
\allowdisplaybreaks
\begin{small}
\begin{equation}
    N_{\mathrm{tab,i}} = \lfloor  D_{\mathrm{cell}} \cdot \frac{N_{\mathrm{tab,benchmark}}}{D_{\mathrm{cell,benchmark}}} \rfloor,
\end{equation}
\end{small}%
\endgroup
where $N_{\mathrm{tab,benchmark}}$ is the number of tabs defined by the manufacturing method, and $D_{\mathrm{cell,benchmark}}$ is the diameter of the cell for which the number of tabs is validated. Here $L_{\mathrm{w}}$ denotes the total winding length. The tab resistance is calculated from the tab geometry and material properties. The resistance of a single tab is defined as
\par\nobreak\vspace{-5pt}
\begingroup
\allowdisplaybreaks
\begin{small}
\begin{equation}
R_{\mathrm{tab,single}} =
\frac{L_{\mathrm{tab}}}
{\rho_{\mathrm{tab}} \cdot A_{\mathrm{tab}}}
=
\frac{L_{\mathrm{tab}}}
{\rho_{\mathrm{tab}} \cdot t_{\mathrm{tab}} \cdot w_{\mathrm{tab}}},
\end{equation}
\end{small}%
\endgroup

\noindent where $L_{\mathrm{tab}}$ is the tab length, $\rho_{\mathrm{tab}}$ is the tab material resistivity, $t_{\mathrm{tab}}$ is the tab thickness, and $w_{\mathrm{tab}}$ is the tab width. For $N_{\mathrm{tab}}$ identical tabs connected in parallel, the equivalent tab resistance of the cell is

\par\nobreak\vspace{-5pt}
\begingroup
\allowdisplaybreaks
\begin{small}
\begin{equation}
R_{\mathrm{tab,cell}}
=
\left(
N_{\mathrm{tab}}
\frac{1}{R_{\mathrm{tab,single}}}
\right)^{-1}
=
\frac{R_{\mathrm{tab,single}}}{N_{\mathrm{tab}}}.
\label{eq: tab_res}
\end{equation}
\end{small}%
\endgroup

The total cell resistance is computed by combining Eq.~\ref{eq: R_ion}, \ref{eq: R_el}, \ref{eq: R_geom_circ}, and \ref{eq: tab_res}. This results in
\par\nobreak\vspace{-5pt}
\begingroup
\allowdisplaybreaks
\begin{small}
\begin{equation}
\begin{split}
    R_{\mathrm{cell}} = R_{\mathrm{ion,stack}} + R_{\mathrm{el,cell}} + R_{\mathrm{geom,cell}} + R_{\mathrm{tab,cell}}.
    \end{split}
    \label{eq: DCIRcell}
\end{equation}
\end{small}%
\endgroup

\subsection{Single-cell Design-Space Exploration and Sensitivity Analysis}
\label{sec:single_cell_analysis}
The influence of the cell design variables on cell capacity, internal resistance, and the gravimetric and volumetric performance indicators is investigated through a design-space exploration and global sensitivity analysis. The analysis uses the design vector defined in \autoref{eq:design_vector}, subject to the component-wise bounds defined in \autoref{eq:design_variable_bounds}. The corresponding numerical values are listed in \autoref{tab:design_variable_bounds}.

\begin{table}[t]
\centering
\caption{Bounds of the cell design variables.}
\label{tab:design_variable_bounds}
\begin{tabular}{lccc}
\hline
Design variable & Lower bound & Upper bound & Unit \\
\hline
$H_{\mathrm{cell}}$                 & 25   & 100  & $\mathrm{mm}$ \\
$D_{\mathrm{cell}}$                 & 15   & 60   & $\mathrm{mm}$ \\
$\sigma_{\mathrm{act,cat}}$         & 15   & 40   & $\mathrm{mg\,cm^{-2}}$ \\
$\varepsilon_{\mathrm{cat}}$        & 0.25 & 0.40 & $-$ \\
\hline
\end{tabular}
\end{table}

For each sampled design, the cell scaling model is evaluated to determine the reversible cell capacity and internal resistance using \autoref{eq: capacity cell} and \autoref{eq: DCIRcell}, respectively. The corresponding performance indicators are defined as

\begin{equation}
    f_{\mathrm{capacity}}(\mathbf{x})
    =
    Q_{\mathrm{rev}}(\mathbf{x}),
\end{equation}

and

\begin{equation}
    f_{\mathrm{resistance}}(\mathbf{x})
    =
    R_{\mathrm{cell}}(\mathbf{x}).
\end{equation}

The gravimetric performance indicator is calculated as

\begin{equation}
    f_{\mathrm{gravimetric}}(\mathbf{x})
    =
    \frac{
        Q_{\mathrm{rev}}(\mathbf{x})
    }{
        W_{\mathrm{cell}}(\mathbf{x})
    },
\end{equation}

where $W_{\mathrm{cell}}$ denotes the cell mass. The volumetric performance
indicator is calculated as

\begin{equation}
    f_{\mathrm{volumetric}}(\mathbf{x})
    =
    \frac{
        Q_{\mathrm{rev}}(\mathbf{x})
    }{
        \frac{\pi}{4}
    D_{\mathrm{cell}}^{2}
    H_{\mathrm{cell}}
    }.
\end{equation}

Two independent sample matrices, $\mathbf{A}$ and $\mathbf{B}$, are generated, each with dimensions $N \times k$, where $N=8192$ is the number of samples and $k=4$ is the number of design variables. Each column corresponds to one design variable, and each row represents one sampled cell design. The entries are sampled independently from uniform distributions within the bounds listed in \autoref{tab:design_variable_bounds}.

The base sample size was increased from $N=4096$ to $N=8192$, which reduced finite-sample fluctuations in the estimated Sobol indices while keeping the computational cost manageable.
For each design variable $x_i$, a hybrid sample matrix $\mathbf{A}_{B}^{(i)}$ is constructed by replacing the $i$th column of $\mathbf{A}$ with the corresponding column of $\mathbf{B}$. The scaling model is subsequently evaluated for each hybrid matrix. The first-order and total-order Sobol indices are estimated using the difference-based Jansen estimators \cite{JANSEN199935}:

\begin{equation}
    S_i
    =
    1-
    \frac{
        \frac{1}{N}
        \sum_{j=1}^{N}
        \left(
            Y_{B,j}
            -
            Y_{A_B^{(i)},j}
        \right)^2
    }{
        2V_Y
    },
\end{equation}

and

\begin{equation}
    S_{T_i}
    =
    \frac{
        \frac{1}{N}
        \sum_{j=1}^{N}
        \left(
            Y_{A,j}
            -
            Y_{A_B^{(i)},j}
        \right)^2
    }{
        2V_Y
    },
\end{equation}

where $Y_{A,j}$, $Y_{B,j}$, and $Y_{A_B^{(i)},j}$ denote the model output for sample $j$ obtained from $\mathbf{A}$, $\mathbf{B}$, and $\mathbf{A}_B^{(i)}$, respectively. For each performance indicator, these outputs form vectors of length $N$. The output variance $V_Y$ is estimated from the combined outputs of $\mathbf{A}$ and $\mathbf{B}$. The first-order index $S_i$ quantifies the individual contribution of design variable $x_i$ to the output variance. The total-order index $S_{T_i}$ additionally includes all interaction effects involving that design variable. Small negative first-order estimates may occur due to finite-sample error and are interpreted as negligible sensitivity.

In addition to the variance-based sensitivity indices, Spearman rank correlations $\rho_i$ are calculated using the combined samples from $\mathbf{A}$ and $\mathbf{B}$. The Spearman coefficient indicates the direction and strength of the monotonic relationship between each design variable and performance indicator. For internal resistance, the sign of $\rho_i$ is inverted such that a positive value represents a favourable reduction in resistance.

To identify favourable regions of the design space, the combined samples are ranked according to each performance indicator. The 10\% of samples with the highest capacity, gravimetric performance, and volumetric performance are selected. For internal resistance, the 10\% of samples with the lowest resistance are selected. Each design variable is normalized according to

\begin{equation}
    x_{i,\mathrm{norm}}
    =
    \frac{
        x_i-\underline{x}_i
    }{
        \overline{x}_i-\underline{x}_i
    },
\end{equation}

after which the parameter distributions within the selected samples are visualized using boxplots. For each performance indicator, the best-performing design is identified from the finite set of evaluated samples.

%% file: Sections/Results.tex
\section{Results}
\label{sec: Results}
This section presents the results obtained using the methodology described in Section~\ref{sec: Methodology}. First, the scaling model is validated against available battery-cell data. Subsequently, the results of the single-cell design-space exploration and sensitivity analysis are presented and discussed.

\begin{figure}[ht]
    \centering
    \includegraphics[width=0.8\linewidth]{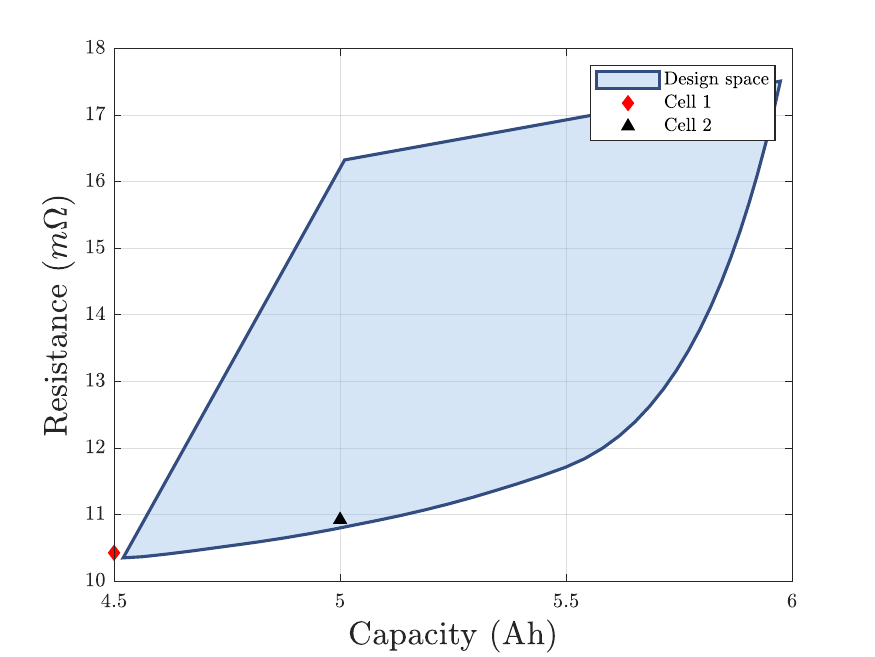}
    \caption{Visualization of the feasible design space with validation cell 1 and validation cell 2, with 2 tabs on the cathode and 3 for the anode. Visualizing the design space of cell 3 with 1 tab for the cathode and one for the anode shifts the predicted design space towards higher internal resistance.}
    \label{fig: capacity range}
\end{figure}

\begin{figure*}[!ht]
    \centering
    \includegraphics[width=0.8\textwidth]{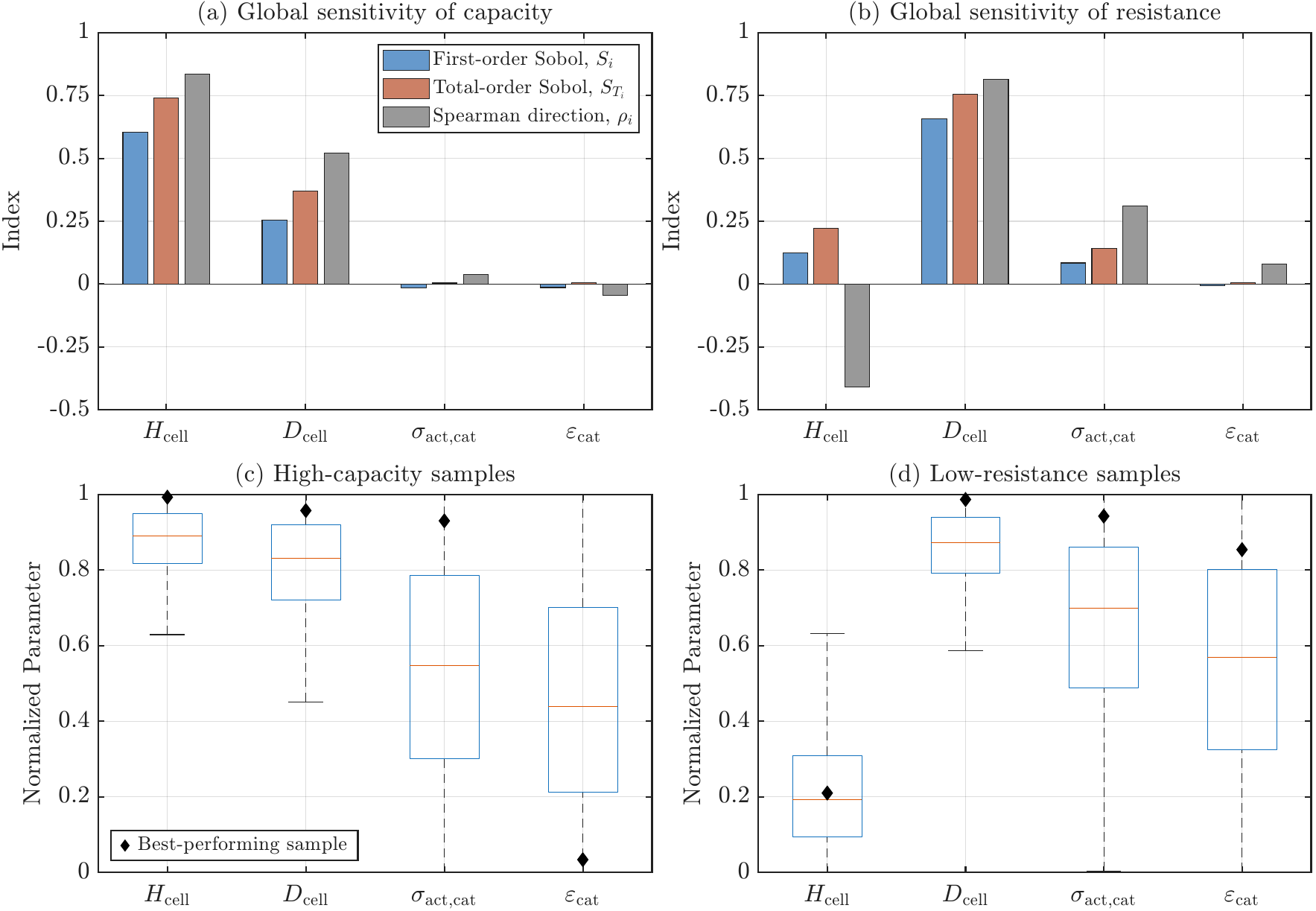}
    \caption{Global sensitivity and parameter distributions for cell capacity and internal resistance. Subplots~(a) and~(b) show the first-order Sobol indices $S_i$, total-order Sobol indices $S_{T_i}$, and Spearman rank correlations $\rho_i$. Subplot~(c) shows, for each normalized design parameter, its distribution within the 10\% of samples with the highest capacity. Subplot~(d) shows the corresponding distributions within the 10\% of samples with the lowest internal resistance. In each boxplot, the horizontal line indicates the median and the box contains the middle 50\% of the parameter values. The whiskers extend to the lowest and highest values within 1.5 times the box height from the lower and upper edges of the box, respectively. Values beyond the whiskers are omitted from the figure. The diamond indicates the parameter values of the best-performing parameter combination for the corresponding performance indicator.}
    \label{fig: Sobol capacity resistance}
\end{figure*}

\begin{table}[b!]
\centering
\caption{Model deviations for benchmark cells}
\label{tab:model_deviation}
\begin{tabularx}{\linewidth}{l >{\centering\arraybackslash}X >{\centering\arraybackslash}X >{\centering\arraybackslash}X}
\hline
Benchmark cell & Capacity deviation (\%) & Resistance deviation (\%) & Winding length deviation (\%) \\ \hline
Cell 1 & 0.6  & 0.7  & 6   \\ 
Cell 2 & 0.09 & 0.1  & 3  \\ 
Cell 3 & 0.8   & 2.1   & 0.6  \\ 
\hline
\end{tabularx}
\end{table}

To obtain the results, a single lithium-ion battery chemistry is considered. The selected chemistry is NMC (Nickel Manganese Cobalt), with NMC-811 used for the cathode, and the anode consists of Silicon-Oxide/Graphite. The chemistry parameters used in the scaling model, linked to this chemistry, are shown in Tab. \ref{tab: cell param} from the appendix. 

\subsection{Model Validation}
The scaling model is validated using cell data obtained from the About Energy database \cite{aboutenergy}, cell data sheets, and literature on battery cell deconstruction \cite{di_prima_calendar_aging_2025,batterymooch_p50b_tabless_2024,molicell_p45b_teardown_2023}. The validation mainly focuses on the resistance and capacity for a fixed battery size (can size). This validates both the geometric scaling and the models used to determine the capacity and resistance. Using benchmark cells that span the predicted capacity range allows the overall consistency of the scaling model to be assessed across different cylindrical cell sizes. The comparison reflects the combined influence of the geometric design variables, electrode properties, and assumed parameters listed in Tab.~\ref{tab: cell param}. In the literature, three cells used in battery model studies and articles show the number of tabs for both the anode and cathode, as well as the winding length, which can be used to validate the model. These validation cells are chosen to span the predicted capacity range. The predicted capacity range and where the actual cells fit in the range are shown in Fig.~\ref{fig: capacity range}. The cells are referred to as cell 1, 2, and 3.
As shown in Eq.~\ref{eq: effective lenght}, there is a strong dependence on the number of tabs used in the jelly-roll. For the specific cells, there are two from the same manufacturer.

When comparing the model's predicted capacity, winding length, and resistance, there is a relationship among the accuracies, since the model predicts some combinations of the design variables that provide approximately the same capacity but different resistance values. The value that matches most closely in both capacity and resistance is selected and then validated for winding length. By comparing the winding length, the model is checked for wall thickness, mandrel diameter, electrode thickness, separator type, and current collector thickness.\\
The first two cells (Cell 1 and 2) have five tabs on the current collectors, two for the cathode and three for the anode \cite{batterymooch_p50b_tabless_2024,molicell_p45b_teardown_2023}.\\
Cell 3 has been deconstructed in greater detail for research \cite{di_prima_calendar_aging_2025}. The research shows a total of three tabs on the current collectors, one for the cathode and two for the anode.\\
The deviations in capacity, resistance, and winding length are presented in Tab.~\ref{tab:model_deviation}. The scaling model predicts the capacity, resistance, and winding length within an acceptable range for the model's goal, with differences arising from assumptions about foil thickness, electrolyte densities, and manufacturing methods. These differ from cell to cell \cite{di_prima_calendar_aging_2025,batterymooch_p50b_tabless_2024}, as shown in Tab. \ref{tab: cell param}, to fit as a rule rather than for a specific cell. Since the overall goal is to optimize the battery design, the manufacturing methods for cells 1 and 2 are adapted to the model. 

\subsection{Single-cell Design-Space Exploration and Sensitivity Results}\label{sec: Single_cell opti}
This section presents the results of the single-cell design-space exploration and sensitivity analysis. The influence of the design variables on the different performance indicators is evaluated to identify dominant design parameters, favourable parameter directions, and relevant trade-offs.

\begin{figure*}[!ht]
    \centering
    \includegraphics[width=0.8\textwidth]{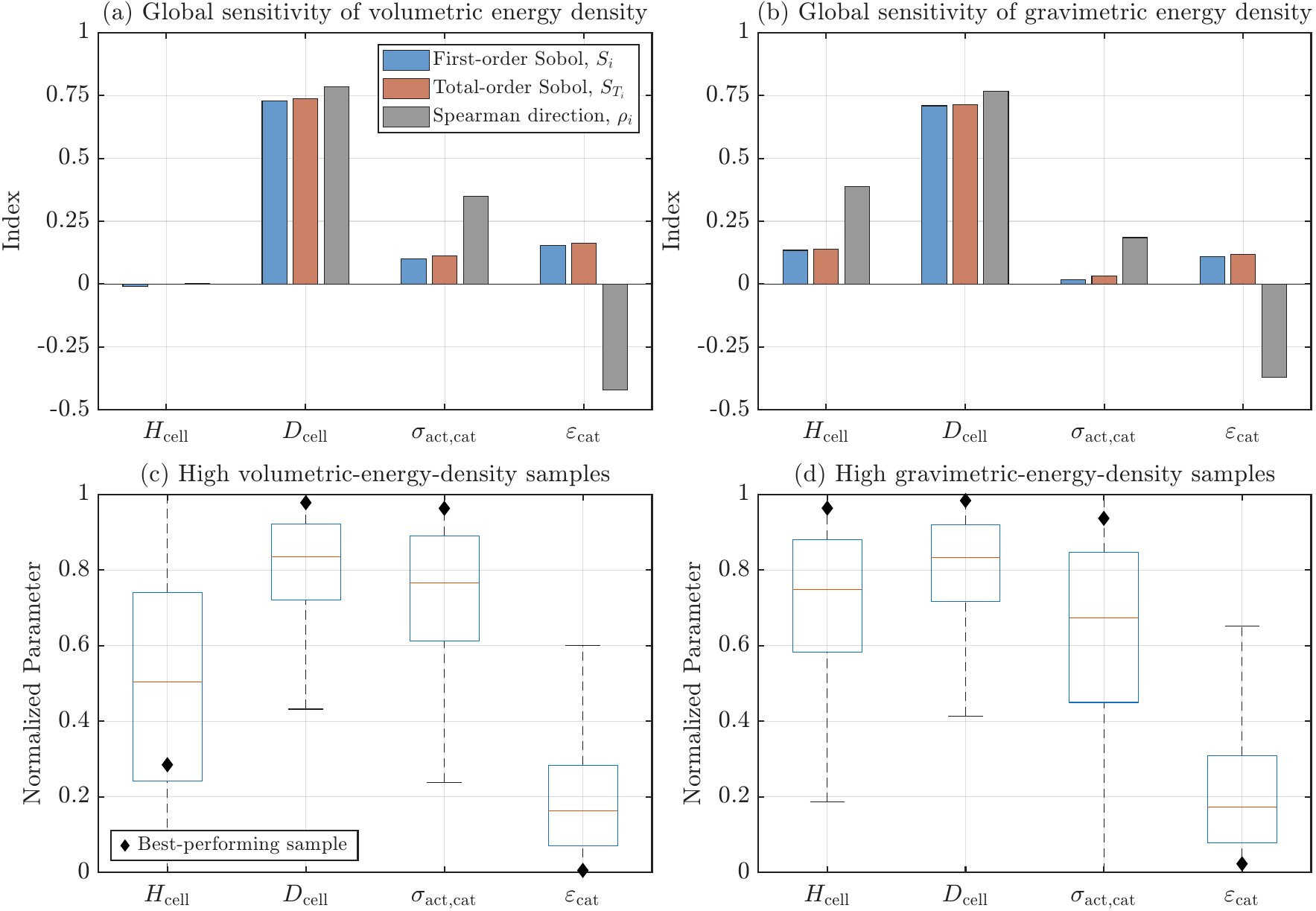}
    \caption{Global sensitivity and parameter distributions for volumetric and gravimetric energy density. Subplots~(a) and~(b) show the first-order Sobol indices $S_i$, total-order Sobol indices $S_{T_i}$, and Spearman rank correlations $\rho_i$. Subplots~(c) and~(d) show, for each normalized design parameter, its distribution within the 10\% of samples with the highest volumetric and gravimetric energy density, respectively. The boxplots are interpreted as described in Fig.~\ref{fig: Sobol capacity resistance}, and diamonds indicate the best-performing sampled designs.}
    \label{fig: Sobol density}
\end{figure*}

First, we evaluate the effects of cathode porosity and active loading on the design space for a fixed volume (can size), as defined in the validation of the scaling model. This design space is visualized in Fig.~\ref{fig: capacity range}. We find that active cathode loading has a greater effect on both resistance and capacity. Regarding resistance, the contributions of all three parts indicate that porosity has minimal impact. The electronic contribution is most influenced by porosity but has the lowest impact on the total resistance. The change in resistance increases with porosity. This is explained by the fact that as porosity increases, paths through the porous medium become more complex, thereby increasing the resistance to following them. For both capacity and resistance, the largest effect is caused by active loading, whose effect on capacity is defined in the active loading definition, which represents the amount of active material per unit area of the current collector. For the resistance, however, an interesting trade-off appears: two parts of the resistance have a counteracting effect. The geometric (resistance of the current collector) decreases with increasing active loading, while the ionic resistance increases with increasing active loading. The effect of active loading explains this: a higher effective loading means a thicker electrode, resulting in a shorter winding length. A shorter winding length decreases the resistance in the current collector (see Eq.~\ref{eq: R_geom_circ}), where in its turn, a thicker layer of active material increases the ionic resistance (see Eq.~\ref{eq: R_ion}).

For the full design space, the effects of the design variables are evaluated using the Sobol sensitivity indices, Spearman rank correlations, and parameter distributions shown in Figs.~\ref{fig: Sobol capacity resistance} and~\ref{fig: Sobol density}.

Overall, the combined sensitivity measures show that cell diameter is the dominant design variable for most performance indicators and generally has a favourable monotonic effect. The preferred cell height depends on the required balance between capacity, energy density, and internal resistance, confirming that no single cell geometry is optimal for all performance indicators. For capacity and internal resistance, the total-order Sobol indices are visibly higher than the corresponding first-order indices for several parameters, indicating the presence of interaction effects. However, the differences remain relatively small, showing that the first-order effects dominate and that interactions provide only a secondary contribution. For the gravimetric and volumetric performance indicators, the response is governed predominantly by direct parameter effects. The distributions of the best-performing samples further show that a relatively large diameter is consistently favourable, whereas cathode active loading and porosity allow a wider range of near-optimal designs and can be adjusted to further refine the gravimetric and volumetric performance. These findings are specific to the selected parameter bounds, uniform and independent input distributions, and imposed relation between anode and cathode porosity.

%% file: Sections/Conclusion.tex
\section{Conclusion}\label{sec: conclusion}
This paper presented a computationally lightweight scaling model for cylindrical lithium-ion battery cells. The model links selected geometric and electrode-level design variables to key performance indicators, including capacity, DC internal resistance, mass, volume, and winding length. The model was validated against available cylindrical cell data by comparing the predicted capacity, internal resistance, and winding length for three benchmark cells.
The results show that cell geometry strongly influences the investigated performance indicators. In particular, cell diameter has the largest influence on most indicators and therefore strongly determines the design direction. Increasing the diameter generally reduces the predicted internal resistance, while increasing both diameter and height increases cell capacity. Cell height, however, introduces a trade-off between capacity and internal resistance. Cathode active loading and porosity have a smaller overall influence, but can be used to fine-tune the balance between resistance, gravimetric energy density, and volumetric energy density.
The proposed scaling model is therefore suitable for early-stage design-space exploration of cylindrical battery cells. Future work should focus on extending the model with temperature-dependent behaviour. Furthermore, the model can be integrated into higher-level battery system and vehicle optimization frameworks. Additional extensions could include ageing effects and thermal or mechanical constraints.

%% file: Sections/Acknowledgment.tex
\section{Acknowledgment}
The authors acknowledge the use of AI tools for code debugging, image generation, and checking grammar, spelling, and writing style.

%% file: Sections/Apendix/Appendix.tex
\section{Model Parameters}
\subsection{Battery Model Parameters}
\begin{table}[H]
\begin{threeparttable}
\centering
\caption{Electrochemical cell model parameters (SI units)}
\label{tab: cell param}

\begin{tabularx}{\linewidth}{l X l l}
\hline
Symbol & Variable & Value & Unit \\ \hline

\multicolumn{4}{l}{\textbf{Cell properties}} \\ \hline
$t_{\mathrm{cell}}$      & Cell thickness $^{a}$                & 0.03 & m \\ 
$t_{\mathrm{cap}}$       & Cell cap thickness  $^{b}$           & $4.1\times10^{-4}$ & m \\ 
$H_{\mathrm{headspace}}$ & Cell headspace  $^{b}$               & $1.0\times10^{-4}$ & m \\ 
$\rho_{\mathrm{cell}}$   & Cell material density          & 7840 & kg\,m$^{-3}$ \\ 
$\rho_{\mathrm{comp}}$   & Electrolyte density            & 1200 & kg\,m$^{-3}$ \\ 
$\mathrm{NPR}$           & N/P capacity ratio $^{d}$             & 1.10 & -- \\ 

$N_{\mathrm{tab,an}}$    & Number of anode tabs $^{b}$           & $n_{\mathrm{a}}$ & -- \\ 
$N_{\mathrm{tab,cat}}$   & Number of cathode tabs  $^{b}$        & $n_{\mathrm{c}}$ & -- \\ 
$N_{\mathrm{tab}}$       & Total number of tabs $^{b}$           & $n_{\mathrm{a}} + n_{\mathrm{c}}$ & -- \\ 

$L_{\mathrm{tab}}$       & Tab length $^{b}$                      & $H_{\mathrm{stack}}$ & m \\ 
$t_{\mathrm{tab}}$       & Tab thickness  $^{b}$                 & $1.5\times10^{-4}$ & m \\ 
$w_{\mathrm{tab}}$       & Tab width $^{b}$                      & $7.5\times10^{-3}$ & m \\ 

\hline
\multicolumn{4}{l}{\textbf{Cathode properties}} \\ \hline
$Q_{\mathrm{dis,1,cat}}$ & Cathode 1st discharge capacity $^{e}$ & 691.2 & A\,s\,kg$^{-1}$ \\ 
$Q_{\mathrm{ch,1,cat}}$  & Cathode 1st charge capacity $^{e}$   & 784.8 & A\,s\,kg$^{-1}$ \\ 
$\rho_{\mathrm{cat}}$    & Cathode composite density $^{e}$     & 4200 & kg\,m$^{-3}$ \\ 
$w_{\mathrm{c,act}}$     & Cathode active material fraction$^{e}$ & 0.94 & -- \\ 
$w_{\mathrm{c,cb}}$      & Cathode carbon fraction $^{e}$       & 0.04 & -- \\ 
$w_{\mathrm{c,b}}$       & Cathode binder fraction  $^{e}$      & 0.02 & -- \\ 
$t_{\mathrm{foil,cat}}$  & Cathode foil thickness   $^{b}$      & $1.0\times10^{-5}$ & m \\ 
$\Delta h_{\mathrm{c}}$  & Cathode height offset $^{a}$         & $-1.0\times10^{-3}$ & m \\ 
$\eta_{\mathrm{FC,cat}}$ & Cathode first cycle efficiency $^{e}$& 0.88 & -- \\ 
$\sigma_{\mathrm{c}}$    & Cathode electronic conductivity $^{b}$ & 2.0 & S\,m$^{-1}$ \\ 

\hline
\multicolumn{4}{l}{\textbf{Separator properties}} \\ \hline
$t_{\mathrm{sep}}$           & Separator thickness $^{e}$           & $1.6\times10^{-5}$ & m \\ 
$\varepsilon_{\mathrm{sep}}$ & Separator porosity $^{a}$          & 0.40 & -- \\ 
$\Delta L_{\mathrm{s}}$      & Separator length offset $^{a}$        & $6.44\times10^{-2}$ & m \\ 
$\rho_{\mathrm{sep}}$        & Separator density $^{e}$             & 900 & kg\,m$^{-3}$ \\ 
$\beta$                      & Bruggeman exponent$^{c}$             & 1.5 & -- \\ 

\hline
\multicolumn{4}{l}{\textbf{Anode properties}} \\ \hline
$Q_{\mathrm{dis,1,an}}$  & Anode 1st discharge capacity $^{e}$   & 2628 & A\,s\,kg$^{-1}$ \\ 
$Q_{\mathrm{ch,1,an}}$   & Anode 1st charge capacity $^{e}$     & 2160 & A\,s\,kg$^{-1}$ \\ 
$\rho_{\mathrm{an}}$      & Anode composite density $^{e}$       & 2140 & kg\,m$^{-3}$ \\ 
$w_{\mathrm{a,act}}$     & Anode active material fraction $^{e}$ & 0.94 & -- \\ 
$w_{\mathrm{a,cb}}$      & Anode carbon fraction $^{e}$          & 0.04 & -- \\ 
$w_{\mathrm{a,b}}$       & Anode binder fraction $^{e}$         & 0.03 & -- \\ 
$t_{\mathrm{foil, an}}$  & Anode foil thickness $^{b}$          & $8.0\times10^{-6}$ & m \\ 
$\Delta x_{\mathrm{a}}$  & Anode offset $^{a}$                  & 0 & m \\ 
$\Delta h_{\mathrm{a}}$  & Anode height offset $^{a}$           & $-1.0\times10^{-3}$ & m \\ 
$\eta_{\mathrm{FC,an}}$  & Anode first charge efficiency $^{e}$  & 0.82 & -- \\ 
$\sigma_{\mathrm{a}}$    & Anode electronic conductivity $^{b}$  & 100 & S\,m$^{-1}$ \\ 

\hline
\multicolumn{4}{l}{\textbf{Electrical and transport properties}} \\ \hline
$\kappa_{\mathrm{e}}$    & Electrolyte ionic conductivity $^{b}$ & 1.0 & S\,m$^{-1}$ \\ 
$\rho_{\mathrm{Cu}}$     & Copper resistivity $^{a}$            & $1.7\times10^{-8}$ & $\Omega$\,m \\ 
$\rho_{\mathrm{Al}}$     & Aluminium resistivity $^{a}$         & $2.82\times10^{-8}$ & $\Omega$\,m \\ 
$\sigma_{\mathrm{tab}}$  & Tab conductivity $^{a}$              & $1.85\times10^{5}$ & S\,m$^{-1}$ \\ 

\hline
\end{tabularx}
\begin{tablenotes}
\footnotesize
\item[a] Assumed
\item[b] Estimated
\item[c] From Ref. \cite{doyle_modeling_1993}
\item[d] From Ref. \cite{batus_np_ratio_2025}
\item[e] From Ref. \cite{faraday_cams_2025}
\end{tablenotes}
\end{threeparttable}
\end{table}